\documentclass[preprints,education,article,accept,moreauthors,pdftex]{Definitions/mdpi} 
\firstpage{1} 
\pubvolume{xx}
\issuenum{1}
\articlenumber{5}
\pubyear{2019}
\copyrightyear{2019}
\history{Received: date; Accepted: date; Published: date}

\pdfoutput=1

\Title{Combining surveys and sensors to explore student behaviour}

\Author{Inkeri Kontro $^{1,\dagger,}$* and Mathieu Génois $^{2,3,\dagger}$}

\AuthorNames{Inkeri Kontro and Mathieu Génois}

\address{%
$^{1}$ \quad University of Helsinki, Helsinki, Finland\\
$^{2}$ \quad CNRS, CPT, Aix Marseille Univ, Universit\'e de Toulon, Marseille, France\\
$^{3}$ \quad GESIS, Leibniz Institute for the Social Sciences, K\"oln, Germany}

\corres{Correspondence: inkeri.kontro@helsinki.fi
}

\firstnote{These authors contributed equally to this work.} 

\abstract{Student belongingness is important for successful study paths, and group work forms an important part of modern university physics education. To study the group dynamics of introductory physics students at the University of Helsinki, we collected network data from seven laboratory course sections of approximately 20 students each for seven consecutive weeks. The data was collected via the SocioPatterns platform, and supplemented with students' major subject, year of study and gender. We also collected the Mechanics Baseline Test to measure physics knowledge and the Colorado Learning Attitudes about Science Survey to measure attitudes. We developed metrics for studying the small networks of the laboratory sessions by using connections of the teaching assistant as a constant. In the network, we found both demographically homogeneous and heterogeneous groups that are stable. While some students are consistently loosely connected to their networks, we were not able to identify risk factors. Based on our results, the physics laboratory course is equally successful in building strongly connected groups regardless of student demographics in the sections or the formed small groups. SocioPatterns supplemented with surveys thus provides an opportunity to look into the dynamics of students' social networks.}


\begin{document}

\section{Introduction}

\subsection{Reformed physics laboratory courses}

Laboratory skills are an essential part of university physics curriculum, and recently much attention has been devoted to what laboratory courses should teach to students \cite{wieman2017}. Traditionally, physics laboratory courses have been attempting to deepen physics knowledge by introducing physics concepts in a new setting, and often they have offered little freedom of experimental procedure. These ``cookbook labs'' have been heavily criticized, as they neither build conceptual knowledge, nor expertise in laboratory skills \cite{royuk2003, holmes2018}.

Many curriculums for investigative laboratory activities exist. Some combine lecture and laboratory courses into a single curriculum. For example, the Investigative Science Learning Environment \cite{etkina2006,karelina2007}, Studio Physics \cite{cummings1999} and SCALE-UP \cite{beichner2007} use an investigative, open process for developing both conceptual and experimental skills.

Another approach is to separate the teaching of physics knowledge and laboratory skills completely. The American Association of Physics Teachers (AAPT) has given out guidelines for physics laboratory courses, which emphasize the role of acquired laboratory skills \cite{aapt}. In this approach, the laboratory courses are constructed around the desired laboratory skills, rather than physics content \cite{zwickl2013, kontro2018}. Laboratory skills include e.g. designing experiments, modelling of physical systems and reporting results, but also interpersonal skills such as collaboration \cite{wieman2017, aapt}.

\subsection{The role of collaboration in physics studies}

The role of collaboration in learning is naturally not limited to a learning goal in experimental physics. Collaborative assignments improve learning and induce conceptual change in students. In physics, discussion-based strategies such as peer instruction are popular. Peer discussions increase the learning of concepts \cite{crouch2001,singh2005}. Guided peer reflection increases the use of diagrams and lead to improved learning gains \cite{mason2016}. The use of peer discussions also improve student attrition \cite{lasry2007}.

The quality of the students' discussion is important for individual learning outcomes \cite{boxtel2000}. Students, who have a tool to test a hypothesis and guidance to accommodate divergent views did well both in discussions and individually \cite{schwartz2007}. Peer discussions can even lead to co-construction of knowledge, where students after a discussion can answer a question neither could answer before \cite{singh2005}. Hence, collaboration is important in students' construction of knowledge.

An important question is how to form the groups. Currently, no consensus exists for whether groups should be mixed (heterogeneous) or matched by ability (homogeneous). For example, Webb et al. found that high-achievers performed more uniformly in homogeneous groups, but a better predictor than group composition was the quality of the group functioning \cite{webb2002}. Cheng et al. similarly found that group composition did not matter, whereas the quality of group processes did \cite{cheng2008}. In any case, in mathematics and natural science, the effect seems small \cite{lou1996}. Also results from college level education vary. A recent study Harlow et al. \cite{harlow2016} found no link between group composition and learning gains in introductory physics, but another study in biology classes found that students with low initial knowledge benefited from homogeneous groups \cite{jensen2011}.

Another important factor influencing group work is that when allowed to form freely, subgroups tend to be less diverse than the group they are drawn from \cite{Larson}. Interactions within a group are naturally not always equal. In a computational study Koponen et al. \cite{koponen2018} found that variations in only two interaction potentials (``competivity'' and ``cooperativity'') led to various group structures. In this study, certain interaction potentials led to group members being excluded from the group. The (lack of) diversity in a small-group may be due to apparent factors, such as gender, but also the underlying factors (differences in knowledge) discussed above \cite{Larson}.

However, the effect of collaboration is not limited to learning and constructing knowledge. Social networks can be characterized through measurements of the importance of a node (person) to the network. For example, betweenness centrality is a measure of how many shortest paths between different nodes go through a certain node, and thus how important this node is for contacts. These kinds of network centrality measures can be combined with student information to predict student performance. For example, when examining networks built from student surveys, network centrality measures and grades correlate significantly, and network centrality measures even predict future grades \cite{bruun2013}. Similarly, networks built from survey data seem to predict student retention \cite{forsman2015}. 

However, network analysis in physics largely relies student answers on whom they remember talking to. Monitoring the evolution of a network is time-consuming and difficult, and the method is less suited for monitoring student activity in a single course, such as a laboratory course. While the laboratory could be monitored by video or audio recording or observations, the analysis of this kind of data is very time-consuming. Following students in large courses can be prohibitively costly.

\subsection{Network analysis of social contexts}

Networks have proven to be a very useful framework to study social structures \cite{otte2002}, as they allow to naturally encode the relations (links) between individuals (nodes). By studying the network structure, one can then extract relevant information about the context it models.

One example is the analysis of the position of an individual in a context. In particular, detecting individuals that are \emph{central} can be of high interest: these individuals can constitute bottlenecks for information flows or spreading processes such as rumor or epidemic spreading. There exist many \emph{centrality measures} \cite{valente2008}, but the most used are \emph{degree} (the number of nodes to which one node is connected), the \emph{strength} (the sum of the weights of the links surrounding one node) and the {betweenness centrality} (the fraction of shortest paths passing through one node). Centrality measures also exists for links, with similar definitions, when one wants to assess which connections play a significant role in a network. However, centrality measures are useful only when one considers networks that are large enough, in order to have a range of values that allows to rank the nodes/links. In small, dense networks, all nodes and links are more or less equivalent with respect to these measures.

Another example is the identification of groups in networks, known as \emph{community detection}. Community detection is a vast topic, and many methods exist to identify relevant groups from the structure of a network \cite{fortunato2010}. One of the standard ones is the maximisation of \emph{modularity}\cite{brandes2008}. Again, these methods are usually tailored for sufficiently large networks, and are not useful when dealing with small, dense networks.

Network analysis provides many tools to extract information from social systems. It is a framework that allows to reduce the complexity of the original situation while keeping all its relevant structures. As a conclusion on its usefulness, let us note that while network analysis usually consists in abstracting a social system as a static network that models the connections between agents, a newer framework has been emerging in the last decades, which allows to take into account the dynamics of such structures. The analysis of such \emph{temporal networks} has and will also provide many new insights to social sciences\cite{holme2012,holme2015}.

\subsection{Measuring development of skill and attitudes}

In the past few decades, the development of physics teaching has led to the adoption of new tools. More and more, students are given diagnostic tests which are used to assess their initial knowledge and what they learned in instruction. These diagnostic tests can be focused solely on physics concepts (e.g. the Force Concept Inventory, FCI \cite{fci}) or they can combine concepts and general physics problem solving (e.g. Mechanics Baseline Test, MBT \cite{mbt}).

Conceptual learning, as measured by conceptual tests, does not necessarily correlate with problem solving ability. In fact, the development of these tests was sparked by the notion that beginning physics students hold many common misconceptions in physics, and that instruction is often not able to address these misconceptions and to induce conceptual change in student reasoning \cite{halloun1985}.

In addition to the conceptual knowledge and conceptual gains of physics students, the expert-like attitudes of students have become a more important topic of study, and several instruments have been developed to study the evolution of student attitudes. These instruments relate student attitudes to those of experts. The Colorado Learning Attitudes about Science Survey (CLASS) is an instrument which has questions that relate to physics studies broadly, rather than to individual courses \cite{class}. CLASS consists of 42 statements, which are scored by a five point Likert scale.

These expert-like beliefs are often seen as a desired learning goal as such. They have also been shown to correlate weakly with learning \cite{cahill2018, perkins2004, ding2014}. They also correlate with experiencing high levels of challenge, interest and skill at the same time (optimal learning moments) \cite{Hendolin}. However, the development of expert-like attitudes is not straightforward. Generally, expert-like beliefs decline with instruction \cite{class, slaughter2011}. Exceptions are mainly courses that focus on modeling \cite{madsen2015}. In many cases, students also know what experts think, but they do not agree when it comes to their own learning \cite{gray2008}.

\subsection{Aim of the research}

In this study, we examine the social dynamics of an introductory laboratory course. This laboratory course is a course with higher than average student attrition. A common criticism from introductory physics students is that it can be hard to get to know other students, and the laboratory, with its group activities, is an important place for building up social contacts. However, we do not know what factors influence the development of social contacts in the laboratory.

Our research questions were as follows:
\begin{itemize}
    \item Can we identify risk factors for being exluded out of laboratory practice?
    \item Do students' pre-existing knowledge influence their risk of dropping out of the course?
\end{itemize}

To do this, we built a network of student contacts in the lab course, and collected information on the student demographics: major, year of study and gender. In this paper, we will show that we can use the role of the teaching assistant in the network to identify the groups working together and hence the strongly and weakly connected students. We also show that on this course, students of all studied demographics have equal roles in the network.

 
\section{Materials and Methods}

\subsection{Physics laboratory course at University of Helsinki}

The introductory laboratory course at the University of Helsinki runs concurrent with the lecture courses in introductory physics, and is mandatory for all students who complete at least 25 credits (European Credit Transfer and Accumulation System, ECTS) of physics. The majority of students on this course are first-year physics majors, and the laboratory is one important place to learn to know other students. Six laboratory assignments are completed each term, and each assignment spans 2-3 weeks, giving students freedom to experiment and iterate their measurement set-up. The course is set up in seven sections, with up to 20 students present at the same time. The assignments are done in small-groups of 3-5 students.

In total, the number of students on the laboratory course is around 140-150 in the beginning of the year. The majority of the participants (~60 \%) are physics major students. The rest are physics minor students, mostly other science majors, or non-degree students participating through open university. An important category of students are pre-service teachers who have mathematics as their first and physics as their second subject.

We have identified the laboratory course as a course which students easily drop out of. To see whether we could identify risk factors in the complex social dynamics of laboratory work, we set out to explore the social dynamics by using the SocioPattern platform. We wanted to see both how social networks evolve in an undergraduate laboratory, and whether we can find correlations between demographics, initial knowledge and being excluded out of collaborative learning in the laboratory.

The students form the small-groups during the first week. They are allowed to change sections and to visit other sections, but the recommendation is to work with the same people the whole year. Subgroups in general are less diverse than the group they are drawn from \cite{Larson}, and we assumed this also to be the case for small groups formed by the students themselves during the beginning of their physics studies. Hence, we collected information on major subject, year of study and gender. We wanted to go beyond surface-level diversity measures and also collected a pre-test on mechanics knowledge (MBT,  \cite{mbt}) and the attitude survey CLASS, \cite{class}. MBT was chosen because the incoming students at UH saturate many conceptual surveys in mechanics, including the FCI.

\subsection{The SocioPatterns platform}
The SocioPatterns platform is a tool developed by the SocioPatterns collaboration\footnote{\url{www.sociopatterns.org}} to collect information about human interactions in the physical world in an automated, observer-free way\cite{cattuto2010}. The original goal was to detect transmission routes of airborne diseases. However, it has since been widely used to study patterns in human interactions in order to analyse social phenomena\cite{stehle2011,barrat2013,fournet2014,genois2015,smieszek2016,ozella2018}

\begin{figure}[H]
\centering
\includegraphics[height=.2\textwidth]{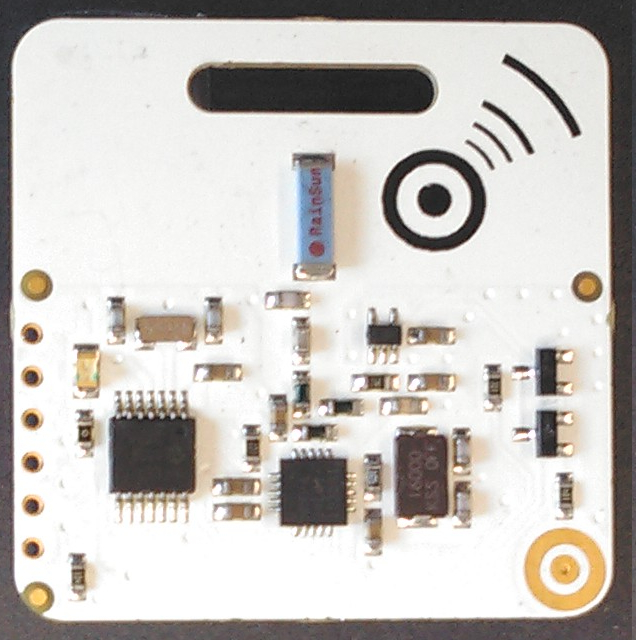}
\includegraphics[height=.2\textwidth]{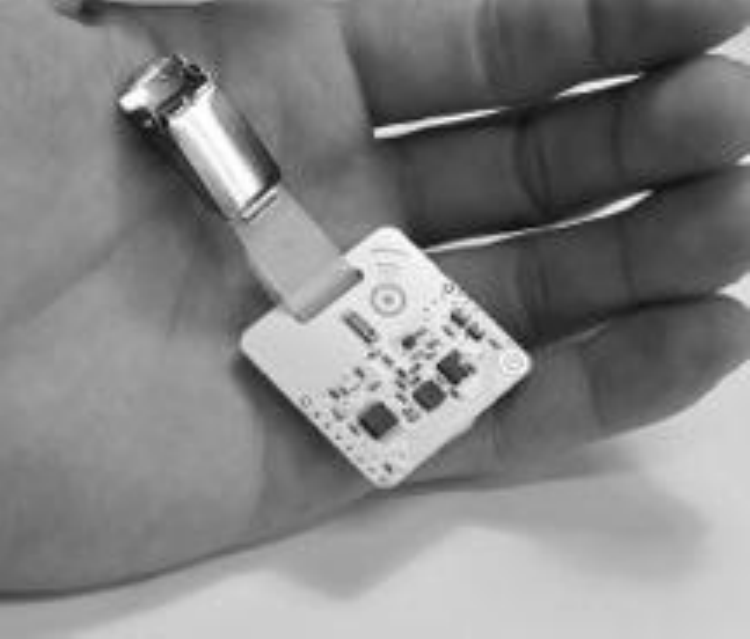}
\caption{\textbf{SocioPatterns sensor.} Front aspect and scale.}
\label{fig:sensor}
\end{figure}

The system is designed to detect face-to-face contacts between individuals. It consists in sensors that are worn by the participants (Figure \ref{fig:sensor}), able to detect each other at short range (1.5 meters maximum) through the use of RFID chips and radio emitters. Furthermore, as the signal used for the detection is blocked by the body, detection is only possible when the individuals are in their respective front half-spheres. A \emph{contact} as detected by this method is thus defined as a physical proximity (less than 1.5 meters), where both individuals are facing each other (they are located in the front space of each other).

Sensors are calibrated so that a contact lasting 20 seconds will be detected with probability $\sim 100$\,\%. Contacts lasting less than 20 seconds are detected with a probability which decreases with their duration. This calibration sets the temporal resolution of the system: contacts are recorded every 20 seconds. The minimum contact duration is therefore set at 20 seconds. The internal detection system of the sensors limits the number of simultaneous contacts to 25 per interval of 20 seconds.

The sensors have a very limited built-in memory: antennas are used to collect and store the contact data. As a consequence, only the areas covered by these antennas are monitored: any contact occurring outside will not be recorded. Antennas have a theoretical range of detection of 30 meters, but it is limited by the presence of obstacles such as walls.

\subsection{The setup}

\begin{figure}[H]
\centering
\includegraphics[height=.3\textwidth]{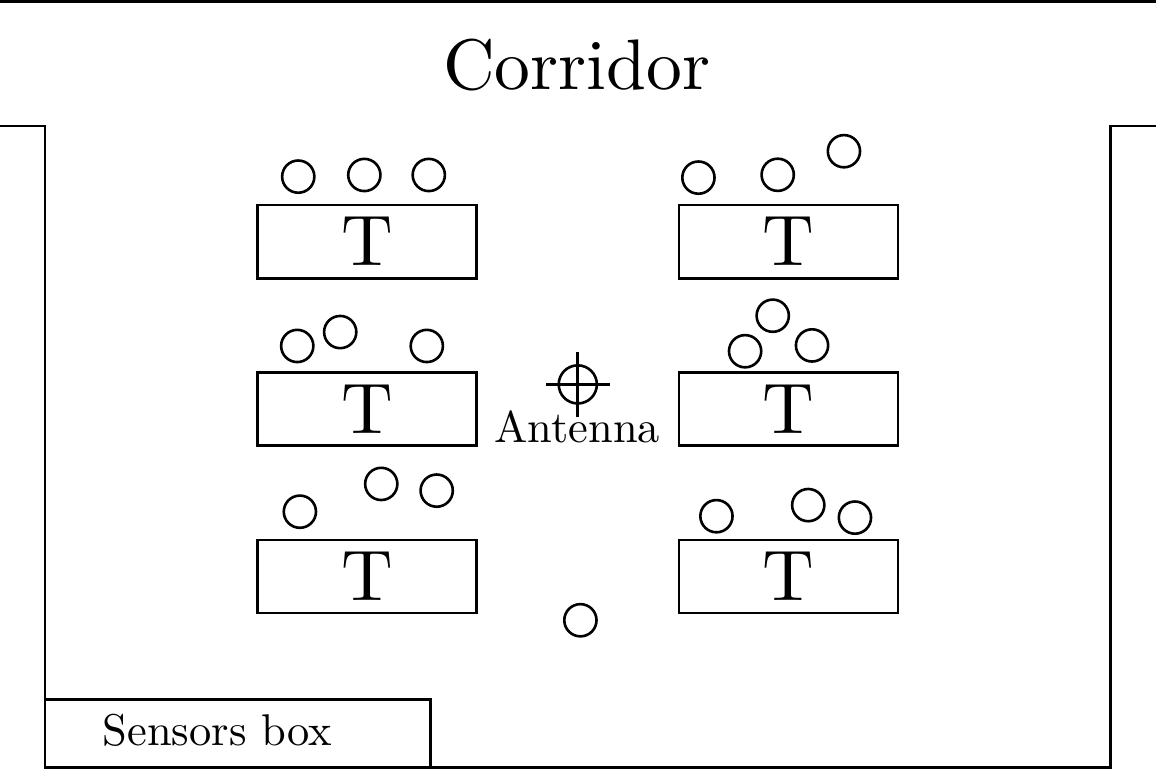}
\caption{\textbf{Setup of the study.} Tables are represented by the rectangles with a capital T. The position of the antenna is signalled by the target symbol.}
\label{fig:setup}
\end{figure}

The studied location consists in a single classroom, intended for practical classes in experimental physics. It consists in two rows of three tables. The room is 11.5 meters long and 9.6 meters wide (Figure \ref{fig:setup}), allowing for the complete location to be monitored by a single antenna.

Students are distributed in seven sections. The class takes place every week on Thursday or Friday depending on the section, and lasts two hours (Table  \ref{tab:groups}). Students are assigned to a section based on their stated preference but may on occasion come to a different one if needed.

\begin{table}[H]
\caption{\textbf{Student sections.}}
\centering
\begin{tabular}{cccc}
\toprule
\multicolumn{2}{c}{\textbf{Thursday}} & \multicolumn{2}{c}{\textbf{Friday}} \\
\midrule
Section 1 & 08:15 to 10:00 & Section 6 & 08:15 to 10:00 \\
Section 2 & 10:15 to 12:00 & Section 7 & 10:15 to 12:00 \\
Section 3 & 12:15 to 14:00 & \\
Section 4 & 14:15 to 16:00 & \\
Section 5 & 16:15 to 18:00 & \\
\bottomrule
\end{tabular}
\label{tab:groups}
\end{table}

Because of equipment limitations, sensors are not assigned to students. Instead, at each beginning of a class students pick a sensor to wear in a box and writes down on a sheet the sensor number along with their student number. At the end of the class, sensors are put back to the box. This setup allows for a better cleaning and separation of the contact data afterwards (see next section). The teaching assistant is assigned to a section, and follows the same procedure. Data collection is started before beginning of the first class of the day and stopped after the end of the last one. It is collected continuously in between.

\subsection{Data cleaning}

Once pre-processed from the raw data, the data collected comes in the form of a $tij$ file, in which each line is a contact occurring at time $t$ between sensors $i$ and $j$ (Figure \ref{fig:tij}).

\begin{figure}[H]
\centering
\begin{minipage}{0.25\textwidth}
\begin{verbatim}
1536214160	1037	1665
1536214160	1037	1052
1536214160	1037	1401
1536214160	1409	1618
1536214160	1409	1674
1536214160	1409	1617
1536214160	1429	1080
1536214160	1429	1422
1536214160	1429	1781
\end{verbatim}
\end{minipage}
\caption{\textbf{Example of a $tij$ file.} This lists 10 contacts, all occurring on time 1536214160. The first line indicates that the contact occurred between sensors 1037 and 1665.}
\label{fig:tij}
\end{figure}   

\begin{figure}[H]
\centering
\includegraphics[height=.2\textwidth]{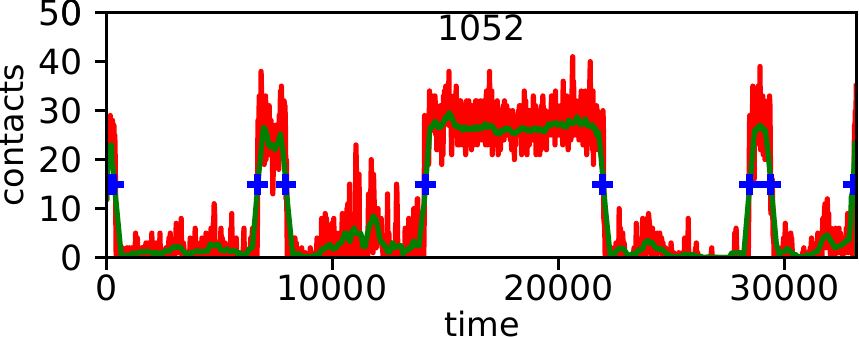}
\caption{\textbf{Example of raw contact activity timeline.} The red line is the raw contact activity ($x$ axis is time in seconds, $y$ axis the number of contacts recorded per time step). The green line is the smoothed contact activity, the blue crosses mark the automatically detected times where the smoothed activity drops below the detection threshold, defining the activity windows of the sensor. In the present example, we see that the sensor 1052 has been used in section 1, 2, 4 or 5, but not in section 3.}
\label{fig:TL}
\end{figure}

Since the same sensor can be used by several students during the day, we use the sheets linking sensor and student number to reconstruct the identity of the students from the data. The method is the following:
\begin{enumerate}
    \item From the sheet with sensor and student numbers, we build for each sensor the list of student who have used it, with their section number.
    \item From the same sheet we extract the list of sensors that were used during the day.
    \item From the data, we extract the raw contact activity timelines of all used sensors. Because we impose the students to put the sensors back in the box at the end of a class, and using the fact the there are always sensors remaining unused in the box, we are able to automatically detect the exact times at which a sensor is taken out of the box and put back in. Indeed, the contact activity drops when the sensor is taken out, as it loses contact with the sensors remaining in the box, and it jumps when the sensor is put back in, as it detects the presence of the sensors in the box. The contact activity is spiky, so in order to improve detection we smooth the curve by averaging on a sliding time window, then we impose a detection threshold. We mark all the times the smoothed curve crosses the threshold: low activity periods thus defined are the activity windows of the sensor (see Figure \ref{fig:TL}).
    \item The automatic detection is not perfect, so we check by hand that all activity windows were detected correctly, and update their definitions where it is necessary.
    \item We compare the detected activity windows to the theoretical ones extracted from the sheet, in order to identify activity windows that might be missing due to sensor malfunctions.
    \item Using the corrected activity windows, we replace the sensor number in the data with an anonymous student ID, remove all contacts between activity windows and involving unused sensors.
\end{enumerate}

At the end of this procedure, we have cleaned data about contacts between students in the same format as Figure \ref{fig:tij}, using anonymous student ID instead of the sensor numbers (a file linking the anonymous ID to the student number is kept, so that we can later link the contact data to the survey data). This is the data we then analyse.

\subsection{Identifying connections}

A network is comprised of nodes and links, which bind the nodes to each other. The contact data forms a temporal network, in which nodes are individual students and links are contacts between them, that appear and disappear as time passes. From this rich and complex data, we compute the aggregated network, in which a link exist between two nodes if they have been in contact at least once, and the weight of this link is the total contact duration between the two individuals.

\begin{figure}[H]
\centering
\includegraphics[height=.7\textwidth]{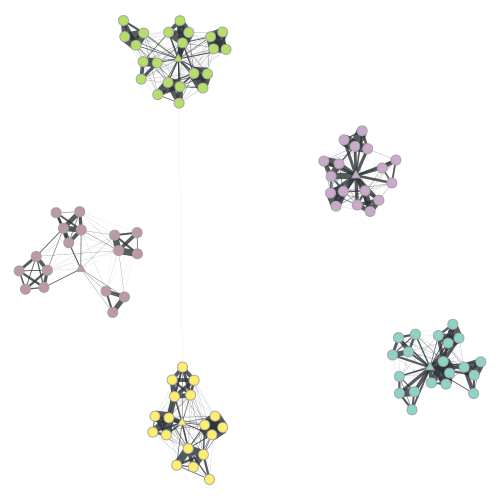}
\caption{\textbf{Example of aggregated contact network.} Colours code the five sections of a Thursday. Round nodes are students, triangle nodes are teaching assistants. The width of a link is proportional to its weight.}
\label{fig:graph}
\end{figure}

As seen on Figure \ref{fig:graph}, teaching assistants (TAs) always adopt a central position in the networks, while students are aggregated in groups around him/her. The next step is identifying which students are working together (forming a small-group).

Usually, the nodes in a network could be characterized by e.g. betweenness centrality, which is a measure of how central a node is to the network. Betweenness centrality for a node is calculated by calculating the shortest paths (number of links weighed by link strength) between all possible pairs of nodes, and calculating the number of shortest paths through each node. Thus, higher betweenness centrality means a more central place in the network. However, the laboratory networks are too small for the betweenness centrality to vary much between students. As only a ~20 people are in the laboratory at the same time, no individuals are far from each other in terms of the number of links needed to get from one person to another. Also, we have numerous contacts between students from different groups. Instead of the usual community detection algorithms, such as the modularity method\cite{brandes2008}, we use a divisive approach, in which we progressively remove links to make the groups appear. 

The traditional method is the Girvan-Newman algorithm \cite{girvan2002}, in which at each step we rank the link in decreasing betweenness centrality and remove the most central one. This algorithm is based on the assumption that a network is made of densely connected groups joined by few links. However, in our case the links between groups are too numerous for such a detection to work. Using the weighted betweenness centrality even worsens the problem, as the strong links exist \emph{within} the groups. Hence, removing links with high betweenness centrality breaks the subgroups, rather than makes them appear.

\begin{figure}[H]
\centering
\includegraphics[height=.7\textwidth]{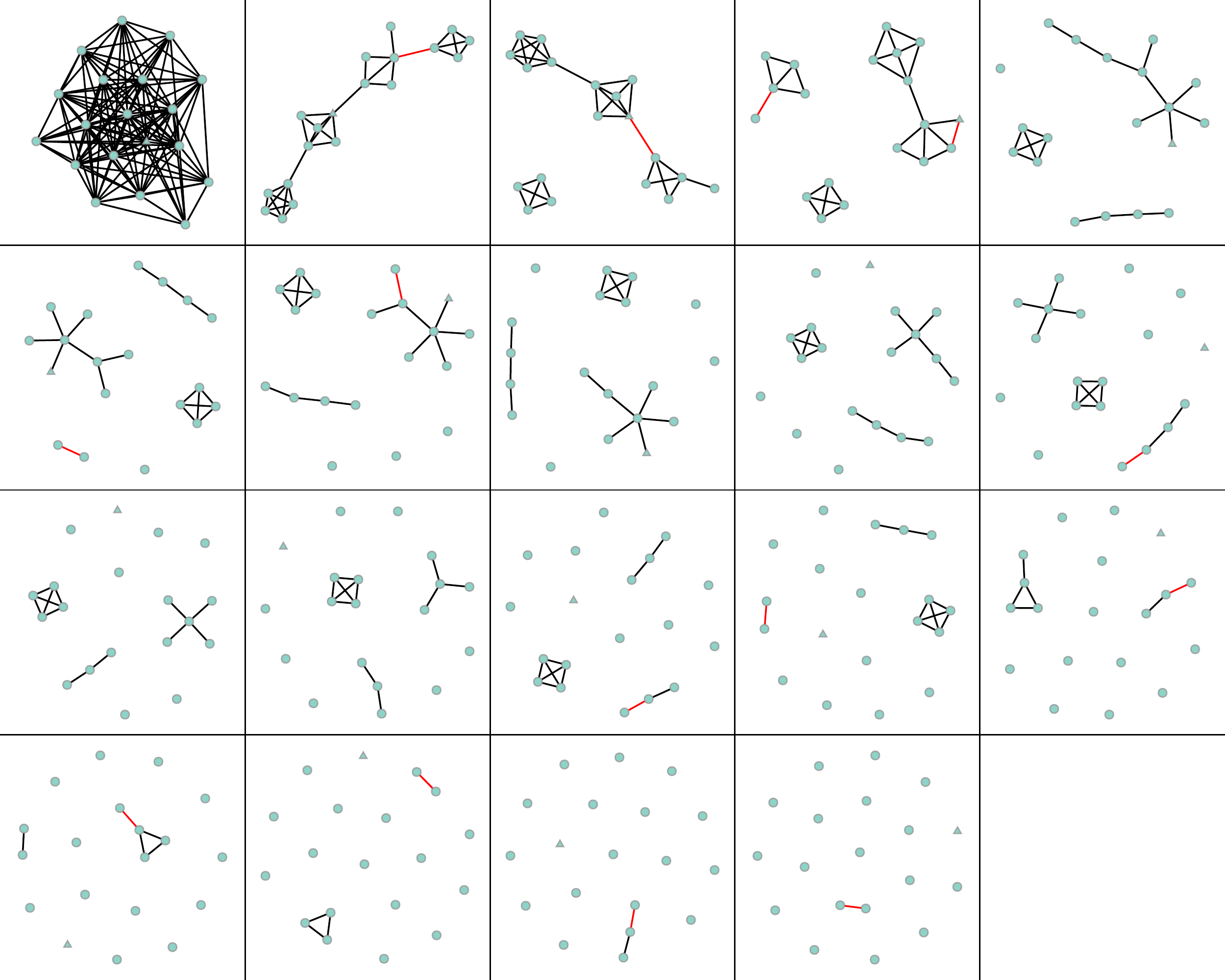}
\caption{\textbf{Depercolation steps for the yellow group from Figure \ref{fig:graph}.} Between each step, links are removed in increasing weights. The steps shown are the one where the remaining network breaks into disconnected parts. The link(s) responsible for the parting is(are) shown in red. The weight of the links are not shown in this representation. Round nodes are students, triangle node is the TA.}
\label{fig:deperc}
\end{figure}

For these reasons, we use directly the weights of the links as the criterion for link removal. Since strong links are within the groups, we remove links starting with the weakest and then following increasing weights. As for the usual method, we then track the \emph{depercolation steps}: the steps in which the number of connected components in the network increases, i.e. the steps where a group breaks away from the network (Figure \ref{fig:deperc}). In the beginning (upper left corner of Figure \ref{fig:deperc}) all links, meaning all contacts between individuals, are present). As the weakest links are removed, progressively, subgroups appear. When stronger and stronger links are removed, the number of subgroups increases, but progressively groups are broken up to individuals as the percolation steps reach a point which removes the strong links inside the groups.

\begin{figure}[H]
\centering
\includegraphics[height=\textwidth]{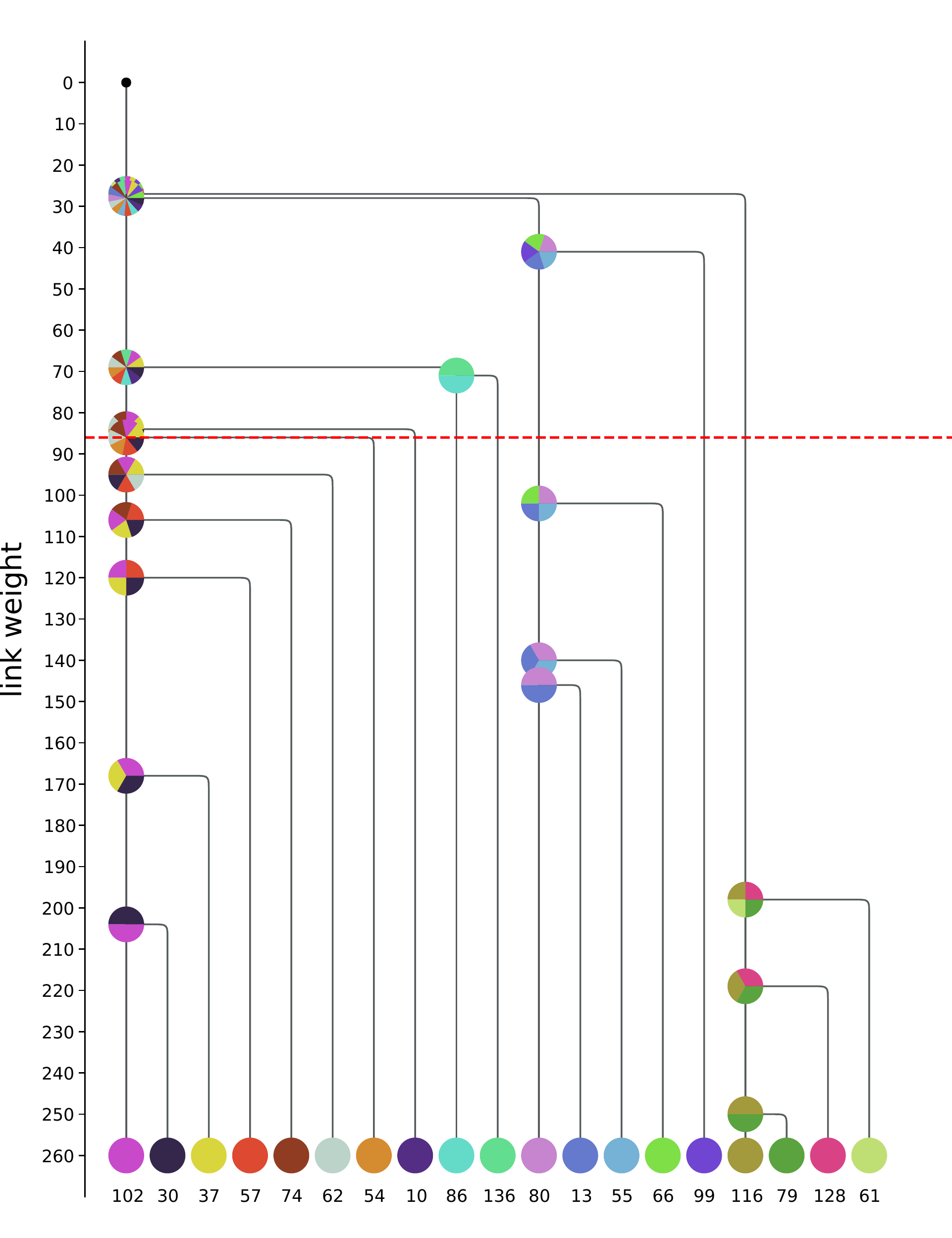}
\caption{\textbf{Depercolation tree for the group of Figure \ref{fig:deperc}.} Branches mark last connections between subgroups as revealed by the depercolation process. The vertical axes represents the weight of the last link connecting two groups, and groups are positioned at the level which they break. The colours of a group code for which nodes constitutes it. The dashed red horizontal line is the TA threshold, \emph{i.e.} the point at which the TA becomes isolated in the depercolation process (in this case individual 54).}
\label{fig:tree}
\end{figure}

We keep track of the groups thus formed at each step in Figure \ref{fig:deperc}, and note the link which removal is responsible for the breaking, along with its weight. This allows us to build a \emph{depercolation tree} joining these groups (Figure \ref{fig:tree}), similar to the dendrogram what one would get from a clustering algorithm\cite{savaresi2002}.

This tree gives the whole internal structure of the network in terms of link depercolation. For example, in Figure \ref{fig:tree}, the group first breaks into a group of four students (61, 128, 79, 116) and the rest of the class. We can thus interpret that this group of four is only weakly connected to the whole group, as all links connecting it to the rest have low weights. However, the students in this group of four are strongly connected to each other.

Such tree gives many possible partitions of the original network, depending on where we cut in terms of limit weight. In our case, we however have a potential natural threshold by using the TA as a reference. At some point in the depercolation process, the TA becomes isolated in the network. By definition, the TA does not belong to any student group. We can thus consider that links weaker than the strongest link connecting the TA to the network are not relevant. In the tree, this is noted as the red dashed line. All groups that break before this point (\emph{i.e.} above the red line in Figure \ref{fig:tree}) are coined as \emph{weakly connected} as they are connected at best with a link weaker than the threshold. All groups that break after (\emph{i.e.} under the red line in Figure \ref{fig:tree}) are coined as \emph{strongly connected} and are assumed to be relevant student groups. In the example of Figure \ref{fig:tree}, that means we have one group of six students (Group A: 102, 30, 37, 57, 74, 62), the TA (54), three isolated students (10, 86, 136), a group of four students (Group B: 80, 13, 55, 66) another isolated student (99) and the previously spotted group of four students (Group C: 61, 128, 79, 116).

The tree also allows us to investigate the inner structure of the groups, by looking at the branching under the threshold line. All of them are made of one strong pair, to which the other nodes are singly connected: (102, 30) for Group A, (80, 13) for Group B, (116, 79) for Group C. Similarly, looking at the branching above the threshold line, we can understand how the groups connect to form the whole network:
\begin{itemize}
    \item isolated node 10 connects to Group A;
    \item isolated nodes 86 and 136 form a pair, which then connects to Group A;
    \item isolated node 99 connects to Group B, which then connects to Group A;
    \item Group C connects last to Group A.
\end{itemize}

This analysis provides thus a method to interpret behaviour from the aggregated network, simplifying the structure by focusing on the strongest links between them. From these connecting steps, one can then make hypotheses about the relations between the students, and the group structure that exists within the class.

\subsection{Evaluating proximity evolution}

The study takes place during several weeks. We can thus explore the way the structures identified through the depercolation method evolve with time.

To do that, we define the proximity between two individuals as the number of steps necessary to go from one to the other in the depercolation tree. The fact that we are dealing with a tree structure ensures that this distance is consistent with the groups as we defined them: two individuals belonging to the same groups will be closer than two individuals from different groups. Having this proximity measure, we can than compute its evolution as time passes.

\begin{figure}[H]
\centering
\includegraphics[width=.95\textwidth]{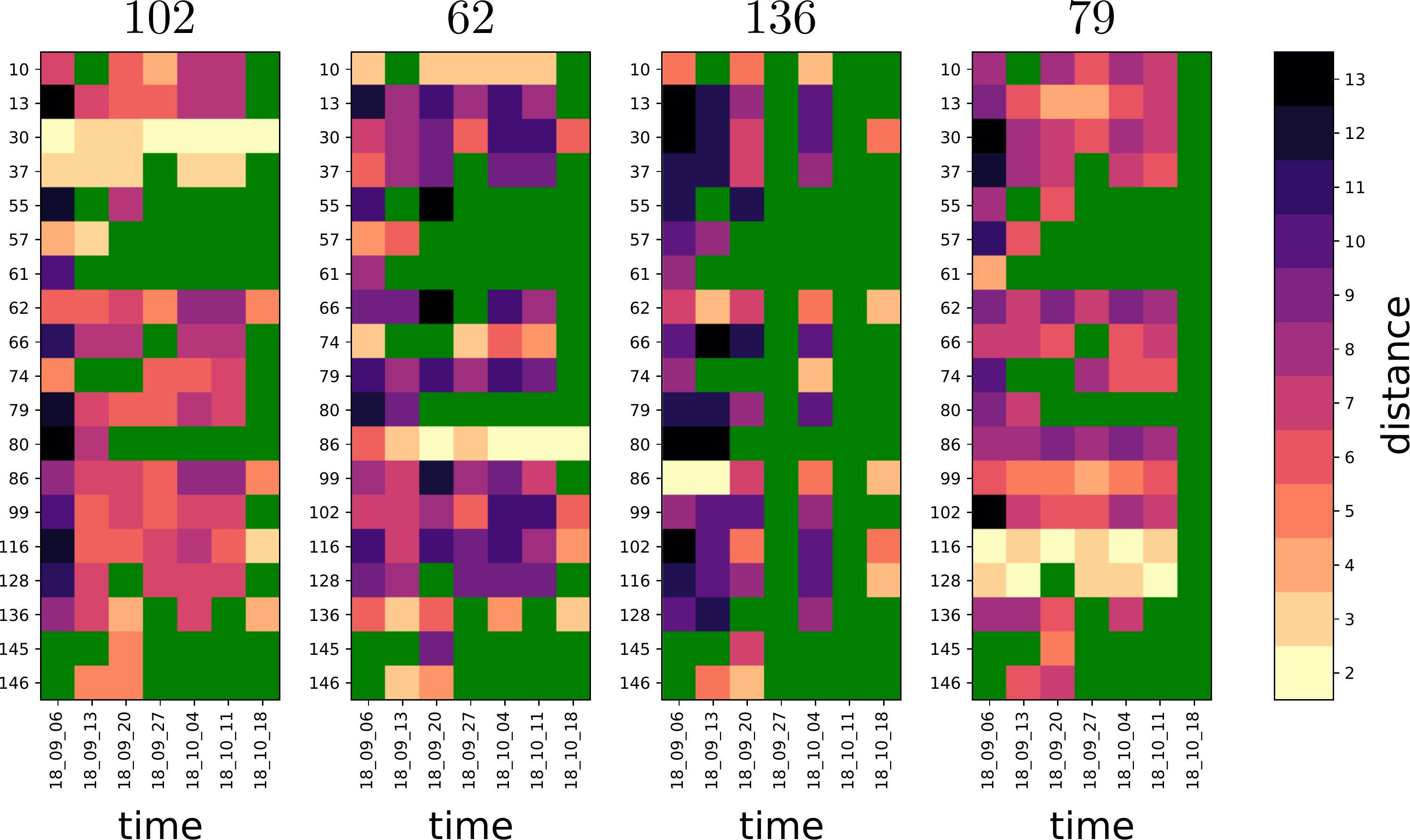}
\caption{\textbf{Evolution of the distance between individuals 102, 62, 136 and 79 and all the other students of their section during seven weeks.} Horizontal axis is time, vertical axis are the students from the section. The colours code for the number of steps separating the two individuals in the tree from Figure \ref{fig:tree} (see colour scale on the right), green codes for missing values due to the absence of either student.}
\label{fig:dist}
\end{figure}

From the examples shown in Figure \ref{fig:dist}, we can see five typical evolution patterns:
\begin{itemize}
    \item stability of a group: whenever the proximity stays high along time, it signals a persistent group, for example the trios (102, 30, 37) and (79, 116, 128), or the pair (62, 10);
    \item stability of separation, whenever the proximity stays low, for example 62 and 13, 62 and 79, 79 and 86, etc;
    \item coming together: when the proximity increases in time, for example 62 and 86;
    \item separation: when the proximity decreases in time, for example 136 and 86;
    \item temporary groups: when groups from occasionally and do not persist, for example 79 and 13 which are close for two weeks in the middle, 136 and 62 which are close only on 13/09 and 18/10, or 79 and 99 which get closer then separate.
\end{itemize}

The results reveal that building on the depercolation trees and the long term monitoring, one can thus have access to the internal dynamics of student groups.

\subsection{Survey data}

All students participating in the laboratory course were asked to complete a background information form, stating their study track (Physical sciences, Mathematics, Science teacher, Chemistry or Other), study year (1, 2, higher or other, where other accounts for non-degree students) and gender (male, female or other), along with their consent to combine these data with data collected from sensors and surveys collected on introductory physics courses. Students who did not wish to participate were instructed to not wear a sensor in the laboratory and to decline their consent in the background survey. Filling in the consent form was compulsory for unlocking the return of lab reports, meaning all students who returned laboratory work for grading had to give or decline consent. Hence, voluntary participation and informed consent were ensured. All data was treated anonymously. As the research also did not involve intervention in the physical integrity of the participants, deviation from informed consent, studying children under the age of 15, exposure to exceptionally strong stimuli, causing long-term mental harm beyond the risks of daily life, or risks to the security of the participants, the study did not require an ethics review, according to the guidelines of Finnish Advisory Board on Research Integrity \cite{tenk}.

To collect information on the students' level of physics knowledge and their attitudes towards learning science, the MBT and CLASS surveys were administered on the physics course lectured concurrently with the laboratory course. The MBT and CLASS surveys are administered as a part of homework exercises, and participation is rewarded with exercise credit equal to one homework problem. This credit was available to students regardless of whether they gave consent to use the data for research. The data were collected electronically and improper data were discarded. For MBT, using less than 300 seconds and for CLASS less than 250 seconds for finishing the survey were used as cut-off. For CLASS, also having more than 4 missing answers, having more than 26 same answers (out of 41), and an incorrect answer to the control item (31) were used to discard improper data. 

The MBT was administered in the first week and CLASS in the second week of studies, which also corresponds to the first and second week of the lab course. The students are meant to take the laboratory and the lecture course at the same time, but many students postpone the laboratory course, and these surveys were not compulsory. Out of the students consenting to participate in the study, 79\% answered the MBT and 75\% the CLASS. 

\section{Results}

The final data set includes 151 individuals, of which 144 are students and 7 are TAs. 13 of the students did not return the consent form, meaning they are excluded from further analysis. The number of students for whom also both CLASS and MBT scores were available was 91.

\subsection{Student participation according to demographics}

We calculated the number of times each student was present in the laboratory and the number of times they were strongly connected, i.e. connected to at least one other student after the percolation step where the TA was detached from the group structure. From this, we calculated the strongly connected percentage (Table \ref{tab:present}). The numbers are strikingly similar for all student groups: men and women, physics majors and other majors, and first year students and non-first year students. Non-first year students were pooled in analysis due to their small number.

\begin{table}[H]
\caption{\textbf{Student participation according to student type.} Average and standard error of the mean of times present, times in a strongly connected group and the percentage of strongly connected groups.}
\centering
\begin{tabular}{ccccc}
\toprule															
Type	&	N	&	Present			&	Strongly connected			&	\% Strongly connected			\\
\midrule															
All	&	131	&$	5.02	\pm 	0.12	$&$	3.72	\pm 	0.13	$&$	75	\pm 	2	$\\
Men	&	76	&$	4.93	\pm 	0.17	$&$	3.64	\pm 	0.18	$&$	75	\pm 	3	$\\
Women	&	55	&$	5.13	\pm 	0.18	$&$	3.82	\pm 	0.19	$&$	75	\pm 	3	$\\
Phys	&	79	&$	5.06	\pm 	0.17	$&$	3.87	\pm 	0.18	$&$	76	\pm 	3	$\\
Other	&	52	&$	4.94	\pm 	0.17	$&$	3.48	\pm 	0.19	$&$	73	\pm 	4	$\\
Yr 1	&	102	&$	4.99	\pm 	0.14	$&$	3.74	\pm 	0.15	$&$	75	\pm 	2	$\\
Yr 2+	&	29	&$	5.10	\pm 	0.24	$&$	3.66	\pm 	0.26	$&$	74	\pm 	5	$\\

\bottomrule
\end{tabular}
\label{tab:present}
\end{table}

\subsection{MBT and CLASS results}

On average, students obtained 15.40  points (59\%) in the MBT. Students majoring in physics had a slightly higher average than non-majors, and non-first year students had higher scores than first-year students (Table \ref{tab:mbt}). There is a statistically significant difference between the scores of men and women ($p=0.001$) and physics and other majors ($p=0.02$, Student's t-test assuming unequal variances).

To calculate the effect size of the difference between men and women, we used Cohen's $d$:

\begin{equation}
d = \frac{M_1-M_2}{SD_{pooled}},
\end{equation}
where $M_1$ and $M_2$ are the means and $SD_{pooled}$ is the pooled standard deviation. The pooled standard deviation is calculated from

\begin{equation}
SD_{pooled} = \sqrt{\frac{(n_1-1)SD_1^2+(n_2-1)SD_2^2}{n_1+n_2-2}},
\end{equation}
where $SD_1$ and $SD_2$ are the standard deviations and $n_1$ and $n_2$ the numbers of observation of the samples to be compared. The effect size for MBT between men and women is $d = 0.56$ and for physics majors and other majors 0.45. This indicates a moderate effect.

The percentage of expert-like answers in CLASS was on average 69.7\% (Table \ref{tab:mbt}). There were no statistically significant differences between women and men, first-year and other students, but the difference between physics and other students is statistically significant ($p=0.013$, Student's t-test assuming unequal variances) with an effect size of $d = 0.56$ (moderate effect).

\begin{table}[H]
\caption{\textbf{MBT and CLASS scores by demographics.}}
\centering
\begin{tabular}{cccc}
\toprule												
Type	&	N	&	MBT			&	CLASS				\\
\midrule												
All	&	91	&$	15.40	\pm 	0.47	$&$	69.7	\pm 	1.6	$	\\
Men	&	52	&$	16.44	\pm 	0.66	$&$	70.3	\pm 	2.2	$	\\
Women	&	39	&$	14.00	\pm 	0.60	$&$	68.9	\pm 	2.3	$	\\
Phys	&	51	&$	16.27	\pm 	0.63	$&$	73.3	\pm 	1.8	$	\\
Other	&	40	&$	14.28	\pm 	0.69	$&$	65.1	\pm 	2.7	$	\\
Yr 1	&	77	&$	15.28	\pm 	0.52	$&$	69.5	\pm 	1.7	$	\\
Yr 2+	&	14	&$	16.14	\pm 	1.19	$&$	70.2	\pm 	4.6	$	\\

\bottomrule
\end{tabular}
\label{tab:mbt}
\end{table}

To study the relationship between the participation metrics (times present, times strongly connected and percentage of times strongly connected to other students) we calculated the correlation coefficients between them, MBT and CLASS. The results are presented in Table \ref{tab:corr}. The correlations are very low for non-dependent variables except for a low to moderate (0.405) correlation between the overall CLASS score and the MBT score. The most likely explanation is that a favourable attitude towards learning physics contributes to good results in a physics problem solving survey, but also that success in physics problem solving contributes to higher self-efficacy, that is beliefs about one's abilities in this context, which likely reflects on the CLASS score.

\begin{table}[H]
\caption{\textbf{Correlation of participation metrics, MBT and CLASS scores.}}
\centering
\begin{tabular}{cccccc}
\toprule												
		&	Present	&	SC	&	\% SC	&	MBT	&	CLASS	\\
\midrule												
	Present	&	1	&		&		&		&		\\
	SC	&	0.589	&	1	&		&		&		\\
	\% SC	&	-0.139	&	0.701	&	1	&		&		\\
	MBT	&	0.063	&	0.059	&	0.014	&	1	&		\\
	CLASS	&	0.106	&	-0.041	&	-0.190	&	0.405	&	1	\\

\bottomrule
\end{tabular}
\label{tab:corr}
\end{table}

We also wanted to see whether students who were strongly connected in the beginning of the course were more likely to participate at the end of the course. For this, we also calculated correlations between times present and times strongly connected in the first four weeks (first two laboratory exercises) of the course to those of the remaining weeks (Table \ref{tab:corr2}). Again, the calculated correlations were small, meaning neither the times the student was present early in the course nor the fraction of times they were strongly connected to other students predicted participation during the last laboratory exercise. Results were similar when only the first two weeks were included in the calculation.

\begin{table}[H]
\caption{\textbf{Correlation of participation metrics for beginning and end of the half-semester. SC = strongly connected.}}
\centering
\begin{tabular}{cccc}
\toprule								
		&	Present beg.	&	SC beg.	&	Present end	\\
\midrule								
	Present beg.	&	1	&		&		\\
	SC beg.	&	-0.12	&	1	&		\\
	Present end	&	0.28	&	-0.10	&	1	\\

\bottomrule
\end{tabular}
\label{tab:corr2}
\end{table}

\subsection{Learning}

As a measure of learning, we collected the grades for the first laboratory assignment. Only 84 students had submitted their report, and for 65 of these students, complete data (including MBT and CLASS) were obtained. This reflects the known problem in the laboratory course: the high drop-out rate. 30\% of students did not receive a grade for this assignment, and hence, did not pass the course during this year.

The first laboratory assignment is to measure the mass of clump of clay without the use of an (existing) scale. Solutions that involve building of e.g. balance scales are allowed. This is an open-ended assignment, meaning that the result (mass of clay) is unknown, and that there are a variety of ways to get an estimate. The students in each small-group need to agree on a common experiment and do the measurements together. The TA answers their questions and helps them, if they are stuck, but the TAs are instructed to not give solutions, but to use e.g. Socrative questioning to help students along. Hence, it is difficult or impossible to perform the laboratory assignment without group work, but it is of course possible for students to engage more or less in their group's experiment. The other experiments are similarly open.

The work is graded through a grading rubric, and for the first assignment, only the results, the (graphical) presentation of them and the uncertainty estimates affect the grade. The learning objectives are experiment design, presentation and rudimentary error analysis.

The students who turned in their work are by most measures equal to those who did not turn in their work. Surprisingly, the students with a grade have slightly lower MBT scores than those without ($15.1 \pm 0.6$ and $16.3 \pm 0.9$, respectively) but this difference is not statistically significant. Neither is the difference in CLASS scores, where students with a grade also have a lower average than those without ($(69 \pm 2)\%$ and $(70 \pm 2)\%$, respectively). 

Students with and without grades are similar also in terms of attendance measures (Table \ref{tab:grade}). 

\begin{table}[H]
\caption{\textbf{Participation metrics for students who received or did not receive a grade for the first assignment. No students received a failing grade.}}
\centering
\begin{tabular}{ccccc}
\toprule															
Type	&	N	&	Present			&	SC			&	\% SC \\
\midrule															
Passed	&	65	&$	5.26	\pm 	0.16	$&$	4.00	\pm 	0.18	$&$	77	\pm 	3	$\\
No grade	&	28	&$	5.18	\pm 	0.20	$&$	3.71	\pm 	0.28	$&$	73	\pm 	5	$\\

\bottomrule
\end{tabular}
\label{tab:grade}
\end{table}

We also calculated the correlations of the participation metrics, MBT, CLASS and grades for the first assignment (Table \ref{tab:corrgrade}.) With the exception of the low correlation between MBT and grade ($r=0.301, r^2 = 0.09$) and MBT and CLASS ($r=0.340, r^2 = 0.12$), the correlations between non-dependent factors are insignificant.
\begin{table}[H]
\caption{\textbf{Correlation of participation metrics at the beginning of the course, MBT, CLASS and grade for the first assignment for 65 students. SC = strongly connected.}}
\centering
\begin{tabular}{ccccccc}
\toprule													
\toprule													
	&	MBT	&	CLASS	&	Present beg.	&	SC beg.	&	\% SC beg.	&	Grade	\\
\midrule													
MBT	&	1	&		&		&		&		&		\\
CLASS	&	0.340	&	1	&		&		&		&		\\
Present beg.	&	0.096	&	0.067	&	1	&		&		&		\\
SC beg.	&	0.111	&	0.000	&	0.577	&	1	&		&		\\
\% SC beg.	&	0.021	&	-0.053	&	-0.005	&	0.754	&	1	&		\\
Grade	&	0.301	&	0.198	&	0.210	&	0.126	&	-0.045	&	1	\\

\bottomrule
\end{tabular}
\label{tab:corrgrade}
\end{table}
\section{Discussion}

\subsection{Factors relating to successful group work}

On average, students participated in $(4.87 \pm	0.12)$ sessions during the seven-week period (70\% participation rate), and were strongly connected to a small-group $(3.60 \pm 0.13)$ times (75\% of times present). As can be seen from Table \ref{tab:present}, students from all demographics participated equally in the laboratory activities. We did not find differences between participation rates of students of different majors, year of study or genders. 

Physics knowledge, as measured by the MBT, or expert-like attitudes in physics, as measured by CLASS, did also not correlate with participation metrics. Additionally, being strongly connected to a small-group during the first two laboratory activities (the first four weeks) did not correlate with participating in the last laboratory exercise (last three weeks). These results are surprising. One would expect that higher physics skills and attitudes towards physics would make the students more committed to coming to class and to work productively. Also, actively attending a class early in term usually makes students more likely to attend later, either because they are more committed from the beginning, because they have already invested time, or a combination of these.

If we cannot determine risk factors from the students who are strongly connected, what about the individuals who are not? Looking at the individual level, two students always percolated away from the tree structure before the TA. An additional seven students were strongly connected only once. These students were evenly distributed across sections, and their demographics followed those of the course at large surprisingly well, considering their small number. For example, five of them (55\%) were first-year students, compared to 56\% of the whole sample, and 3 (33\%) were female, compared to 43\% of the sample.  

Reasons for being excluded out of group work are not naturally limited to neither surface-level nor the deep diversity measures considered here. In a computational study, variations in interaction (``competivity'' and ``cooperativity'') led to the formation of different small group dynamics, including exclusion of group members \cite{koponen2018}. Our experimental set-up does not allow for monitoring of interaction type or direction, but we can see similar patterns: some students are consistently outside the small groups.

We were not able to identify risk factors for being excluded from small-group activities or dropping out of the course from this sample. This is a positive result: the set-up of the laboratory course seems to serve students of different demographics and with different initial skills equally. While we did not see correlations between participation, grades and retention on this single course, on a larger scale, network centrality measures have been used to predict grades and student retention \cite{bruun2013, forsman2015}. Hence, it is important that students participate and cooperate in instruction. The participation rate on our course satisfactory, bearing in mind that a laboratory assignment may be completed in one week instead of the two or three that are allocated to it.

\subsection{Stable groups}

\begin{figure}[H]
\centering
\includegraphics[width=0.5\textwidth]{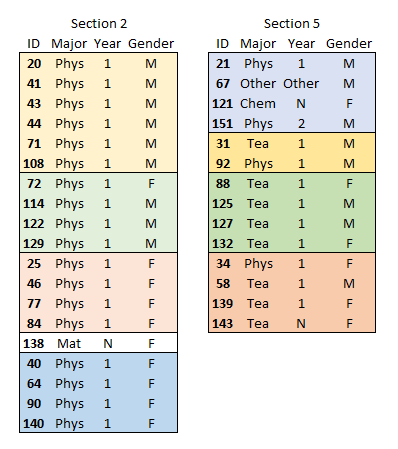}
\caption{\textbf{The strongly connected groups of sections 2 and 5 during the third laboratory week, with demographics (major: Physics (Phys), Teaching (Tea), Mathematics (Mat), Chemistry (Chem), or other, Year: 1, 2, higher (N) or other and gender.)}}
\label{fig:diversity}
\end{figure}

Generally, student groups were less diverse than randomly assigned groups would be. For example, in section 2, a particularly homogeneous section which consisted mainly of physics majors in their first year, the groups split along gender. In week 3, the students formed three strongly connected single-gender groups (two female, one male group) and only one mixed-gender group (Figure \ref{fig:diversity}). One student was loosely connected to the structure.  

On the other hand, in the more diverse section 5, pre-service teachers formed the majority. In week 3, the section had one strongly connected mixed-gender group of first-year teaching students and one single-gender pair of first year students, but the other strongly connected groups were heterogeneous both in terms of major, year of study and gender (Figure \ref{fig:diversity}). While there were fluctuations in group composition over time, we can track most of these groups over the seven-week period. However, new and transient groups form in the data.

The process of forming the groups has random elements. First, students sign up for the sections by listing three in order of preference, and sections are formed by administration, who have no knowledge of friendships or social patterns. Also, as the students sign up to several courses during the same week with similar admission processes without cross-checked admissions, some students may want to change sections once all selections are finalised. 

During the first week, students form groups in the sections by their own accord. They are instructed to work in these groups until the end of the term, but as students' schedules can change and some students drop the course, there are natural fluctuations in the structure of the small-groups. Students may also be absent due to e.g. illness. Groups that end up too small are encouraged by the TA to merge and too large groups to split. Naturally, students tend to seek out groups in which they know other students, which may explain the split by gender in section 2 (Figure \ref{fig:diversity}). This is also expected, bearing in mind that diversity within groups tends to be smaller than diversity between groups \cite{Larson}. The data is very rich and the analysis could go in several directions. However, in this experiment, we have to account for the possibility of missing data. For example, in both sections 2 and 5, in week 3, we observe a group that consists of a majority of male students but also one female student (Figure \ref{fig:diversity}). Physics education research literature has examples of female students not thriving in an otherwise male group \cite{heller1992a,heller1992b}, although this is not always the case \cite{harlow2016}. Indeed, in later weeks, student 72 goes on to be strongly connected to a new, transient group where she is not the only woman. On the other hand, student 121 shows repeatedly to be strongly connected to the same group where the other students are male. Unfortunately, due to the possibility of missing data (students not wearing sensors or students wearing sensors but not consenting to data use) we are not with this data set able to say whether these students definitely are the only women in their respective groups.

Despite the many random factors influencing group composition and dynamics, we can see that many of the groups appear stable already in the first week and remain that way until the end of the observation period. For example, the trio of student IDs 102, 30 and 37 appear close whenever the whole group is present (Figure \ref{fig:dist}) and this can also be seen in their depercolation tree (Figure \ref{fig:deperc}). 

\subsection{Learning outcomes}

The course grading is based on individual laboratory reports. The learning outcomes in terms of grades seem unaffected by whether the student is strongly connected to a group. Also, the subset of students who turned in their work are very similar to students who did not, and even the attendance on the course (during the first seven weeks) does not set them apart. This is unexpected, because we cannot see a correlation between students turning in laboratory work and students attending the laboratory sessions.  Neither the number of times the students attended the laboratory sessions nor the times they were strongly connected to other students seem to influence whether the student will turn in their work for grading. A possible reason for the absence of correlation between times present and learning outcomes is that there are multiple possible reasons for absence. Being away may signal low commitment, but well-performing groups may complete the assignment more quickly and agree to not attend some sessions.

There is a low correlation between mechanics knowledge (MBT) and the grade. The grade does not measure mechanics knowledge directly, but a good knowledge of Newtonian mechanics and problem solving certainly helps in obtaining good and reliable results. Despite this, the correlation coefficient is only $r=0.301$. We used the grade as a proxy for learning, but naturally many things influence the grade. As the grade is awarded a written report, the students' abilities to produce text and figures affect the grade. Also the evaluation of the goodness of the experimental set-up and protocol are viewed through the students' written descriptions of the experiment, rather than the experiment directly. Hence, we cannot from this set of data reliably say much about the relationship between group work and learning. This is a known problem in evaluating student performance in laboratories. Communicating science through a written report is naturally a learning goal of the undergraduate laboratory \cite{aapt}, but it should not be the only learning goal measured. A possible solution to this problem would be the adoption of the working grade \cite{dunnett2019}, which can be used to assess working habits and group work during the laboratory session.

Previous research has shown that CLASS scores correlate with retention on introductory physics courses \cite{perkins2004}. We did not observe this in our data, but our data is from a short segment of a course. In terms of social dynamics in the laboratory, the course seems to function well. All TAs play a similar, central role in the dynamics of their respective sections, and can be used as a treshold to separate the different groups. The number of students who are mostly loosely connected to the network is small, and most students attend the course regularly and while there are fluctuations in the composition of the strongly connected groups, most groups in each section persist and most students are strongly connected to their group in any given laboratory session. Pre-existing physics knowledge or attitudes towards learning science seem to have little impact on the group dynamics or the learning outcomes. In literature, the effects of group composition on group work vary significantly, but a recent study in introductory physics found that group composition had no effect on learning gains \cite{harlow2016}. Clearly, the problems that lead to students dropping the course are outside the laboratory sessions.

\subsection{Limitations of the study}

A network study that combines surveys and sensor data in a authentic setting means that the case of missing data has to be accounted for. In our case, the data was collected in the first weeks of the semester and, indeed for many students, the first weeks of university studies, which means that in the student pool, there are students who only try out physics. Their commitment is low, and they can easily drop courses. Of course, students also have a right to decline to participate in a study. The students were asked not to wear a sensor if they did not participate in the study, but nevertheless we had data from students who either declined consent or did not return the consent form. Further, not all students who gave consent took all surveys. Our solution to this was to only include in the analysis the data from the students who gave informed consent at each stage. This likely skews the analysis towards a more regular student population: students who complete all assignments and students who take courses in the recommended order (as the surveys were collected in the concurrent lecture course).

In demographic terms, the students who consented to participate in the study are very similar to the students who start introductory physics at the University of Helsinki. 60\% are physics majors and 40\% female. The majority are first year students and the absolute majority of the students who are not, do not have physics as a major. The sample also includes some non-degree students, who generally are either students through the open university or teachers, who are working on the qualifications for another subject to teach. While the commitment and diligence of the students who participate may be higher than average, the sample does cover all typical kinds of students on our laboratory courses.

Another limitation that must be addressed is that in our study, we have used face-to-face time as a proxy for interactions. This naturally induces a possibility of bias, because it is not possible to trace the quality of interactions. However, we believe that in the studied setting, detecting face-to-face contacts is a suitable proxy for contacts. The laboratory assignments are open ended, meaning the students have to decide on a common strategy (with or without the help of the TA). It is thus difficult to perform experiments without communication, and as attendance in the sections is not compulsory, students do not attend if they have no experiment to do. However, we cannot know the type of interaction between students, and not all interactions are bound to be beneficial neither for the laboratory assignment nor for social inclusion.

\section{Conclusions}

In this paper, we have shown that the combination of the SocioPatterns platform and supplementary data provides an opportunity to follow the social patterns of students working in small groups on a laboratory course. By asking students for their background information and by combining it with the data obtained for the network analysis, we can study the group composition and evolution of the group dynamics. 

We introduce a procedure in which links are removed from weakest to strongest. The TA is used for calibration: groups that are connected at the depercolation step which detaches the TA are classified as strongly connected. Most students are strongly connected most of the times they attend the laboratory, and the attendance is reasonably high. The advantage of this procedure over conventional network detection algorithms is that we are able to identify small groups in a small, tightly connected network. We can identify typical interaction patterns in the laboratory: stable groups, transient groups, where the proximity changes over time, and stable separations, where certain students do not interact with each other.

Students of different demographics had equal risks of dropping out (measured both by attendance at the end of the observation period and turning in work) or not being strongly connected to a group. We observed that some students were rarely or never strongly connected to a group, but these students did not share any characteristic or set of characteristics. Clearly, for this course, factors influencing dropping out are more subtle than simple demographics. This means that the social dimension of the course works equally well for students of e.g. different major subjects. On the other hand, improving the social dynamics in the laboratory is unlikely to change the drop-out rate.

\vspace{6pt}


\funding{This research received no external funding.}


\conflictsofinterest{The authors declare no conflict of interest.}


\reftitle{References}

\end{document}